# On Divergence in Radiation Fields
**Toward highly focused neuron stimulation for neurology and HCI**

Joerg Fricke
Dept. of Mechanical Engineering
University of Applied Sciences Muenster
Muenster, NRW, Germany
j.fricke@bkbmail.de

ABSTRACT

Three thought experiments demonstrate that under certain circumstances static or radiation fields have to be attenuated or amplified multiplicatively in order not to violate the conservation of energy. Modulation of radiation by means other than superposition is theoretically made possible by plugging additional terms into the source slots of the Maxwell equations. Modulated radiation would enable the well focused stimulation of neurons for diagnostic and therapeutic purposes.

## I. INTRODUCTION

The Maxwell equations describe basic relationships whose validity within the domain of classical electromagnetism is beyond doubt after successful application for one and a half century. However, there are reasons to believe that the question is still not settled which quantities have to be plugged into the "source slots" offered by the equations, starting, as far as I know, with Heaviside's duplex equations. The modern point of view is that charges and currents of charges are inherently coupled to particles, and that they are the sources of fields. But there were attempts to introduce (sometimes implicitly) some kind of "particle-free" field divergence which is not the cause but an effect of the field [1]. For the most general case, the duplex equations could be enhanced to read

$$\nabla \times \mathbf{E} = -\frac{\partial \mathbf{B}}{\partial t} - \rho_m \mathbf{u_m} - \varsigma_m \mathbf{v_m} \qquad (1)$$

$$\nabla \times \mathbf{H} = \frac{\partial \mathbf{D}}{\partial t} + \rho_e \mathbf{u_e} + \varsigma_e \mathbf{v_e} \qquad (2)$$

with $\rho_e$ and $\rho_m$ denoting the particle-bound densities of the divergence of $\mathbf{D}$ resp. $\mathbf{B}$, $\varsigma_e$ and $\varsigma_m$ standing for the particle-free densities of the divergence of $\mathbf{D}$ resp. $\mathbf{B}$, $\mathbf{u_e}$ and $\mathbf{u_m}$ representing the velocities of the particles contributing to $\rho_e$ resp. $\rho_m$, and $\mathbf{v_e}$ and $\mathbf{v_m}$ being virtual velocities assigned to $\varsigma_e$ resp. $\varsigma_m$. Usually, the current density $\mathbf{J_e} = \rho_e \mathbf{u_e}$ is introduced, and accordingly one could define the current densities $\mathbf{J_m} = \rho_m \mathbf{u_m}$, $\mathbf{J_{\varsigma e}} = \varsigma_e \mathbf{v_e}$, and $\mathbf{J_{\varsigma m}} = \varsigma_m \mathbf{v_m}$. Of course, according to the state of the art, $\rho_m = 0$, and

$$\varsigma_e = 0 \qquad (3a) \qquad\qquad \mathbf{J_{\varsigma e}} = \varsigma_e \mathbf{v_e} = \mathbf{0} \qquad (3b)$$

$$\varsigma_m = 0 \qquad (4a) \qquad\qquad \mathbf{J_{\varsigma m}} = \varsigma_m \mathbf{v_m} = \mathbf{0} \qquad (4b)$$

As a consequence of Eq. 3 and 4, fields cannot be attenuated multiplicatively but by superposition only. This is certainly true for static fields. As an illustration for radiation fields, consider two separated volumes of space $V_0$ and $V_1$ surrounded by vacuum, each containing a known distribution of charges and currents causing the fields $R_0$ resp. $R_1$. Assume that $R_0$ in the absence of the sources in $V_1$ is already known. It is legitimate to calculate the net field by first calculating $R_1$, taking into account the sources inside $V_1$ only, and then determining the net field by adding $R_0$ and $R_1$. Both $R_0$ and $R_1$ comply with the Maxwell equations, and so does the net field, of course. Eq. 3 and 4 hold if the power exchanged between $R_0$ and the sources in $V_1$, and between $R_1$ and the sources in $V_0$ is exactly balanced by interference of $R_0$ and $R_1$. This is usually the case but the subject of this paper is the demonstration of counter-examples.

A seeming exception is the concept of "penetration depth". If $V_1$ would contain only a thick sheet of metal, one could multiply $R_0$ inside the metal with a term $\exp(-s/\delta)$ where $s$ denotes the distance from the surface and $\delta$ the penetration depth, calculate the loss, and limit $R_1$ to the reflected wave. While this procedure yields the measurable quantities correctly, each field component in itself does not comply with the Maxwell equations if Eq. 3 and 4 hold. A rigorous treatment has to be based on the extinction theorem [2].

The application of Eq. 3 and 4 to radiation fields is reasonable but not based on direct experimental data [3]. Actually, a direct measurement of the divergence of a radiation field does not seem to be feasable. The aim of this paper is to provide three counter-examples to Eq. 3 and 4 in form of thought experiments. The rest of the paper is organized as follows: In the second section, assumptions and characteristics of the systems employed in the thought experiments are delineated. The third section describes the experiments itself, while the fourth one outlines the immediate consequence for the theory and a way of experimental verification. The appendix provides the details of the calculations done in the framework of the first and second thought experiment.

## II. METHOD

The direct method of testing the general validity of Eq. 3 and 4 were to consider a charged particle moving in a radiation field as fundamental as possible. The energy spent by the radiation source has to equal the sum of the energies absorbed by the charge and radiated to infinity [4]. However, this method turns out to be difficult for several reasons:

- It seems to be impossible to solve the problem in a symbolical way; therefore one has to resort to numerical calculations.
- Generally, the phase relation between the incident field and the field scattered from a point charge on a closed surface at a large distance from the particle is a highly oscillatory function of position. Therefore, numerically integrating the power flow through such a surface is error prone resp. computationally costly.
- Arbitrary fields can be calculated by integrating over an angular spectrum of plane waves. So, inhomogeneous plane waves can be considered to be among the most elementary components of fields, with uniform plane waves being a limiting case. However, there can be no pure inhomogeneous wave without any material boundary. It follows that the treatment has to take into account the interaction of charged particle and boundary.

The second and third difficulty does not occur if the point charge is replaced by a nonradiating system. It is well known that if Eq. 3 and 4 hold a nonradiating system cannot absorb energy from radiation fields. So, in order to disprove the general validity of Eq. 3 and 4 it is sufficient to demonstrate a nonradiating system absorbing energy. Obviously, this is a case where dealing with a system is easier than treating its components separately.

The only requirement for a thought experiment is that it does not violate any laws of physics. Technical feasibility or the probability of occurrence of the situation under consideration in a "natural" context are inessential. The following features greatly facilitate the calculations without introducing any basic impossibilities:

- Each (macroscopic) component of the system moves on predetermined paths not influenced by the incident radiation field. Every detail of the incident field is known in advance, so this feature could be achieved by exchange of energy and momentum with a remote system without any feedback. For example, the frequency of the incident field could be located in the microwave domain while the exchange is carried out via visible light.
- The system does not scatter the incident field but is completely transparent. This could be accomplished by dividing the system into electrically small parts which are coupled by controlled current sources, i. e. the system is composed of an active controlled metamaterial. As above, no feedback mechanism is required because the incident field is known in advance.
- All actions, whether mechanical or electrical, are performed with an efficiency of 1.

The first feature has two consequences: Regardless of the force exerted by the incident field, the center of the nonradiating system can move with constant velocity, avoiding difficulties related to mass or self-force which could be encountered when dealing with accelerated particles at relativistic speed. If the incident field loses or gains energy, the difference of this energy and the growth or loss of internal energy (e. g. internal stress) is transferred via the nonradiating system ultimately to or from the remote system.

## III. THOUGHT EXPERIMENTS

*A. Spherical Shell with Constant Radius in an Inhomogeneous Field*

The magnitude of the phase velocity $\mathbf{u_P}$ of inhomogeneous plane waves is less than the speed of light; so, in principle, any material object can move with such a field in a phase-locked way. Here a monochromatic field is employed.

In this first thought experiment, a nonconducting, evenly charged spherical shell maintains a constant radius. At its center, a point charge is located of opposite polarity but same amount as compared to the charge on the shell. Consequently, there is no field outside the shell. The shell moves uniformly with a speed $u_S$ not much less than the speed of light, e. g. $u_S = 0.9\,c$ which yields $\gamma = 1/\sqrt{1-u_S^2/c^2} \approx 2.29$. So, seen from the laboratory rest frame (LRF), the shell is distinctly spheroidal. In Cartesian coordinates, the center of the shell moves parallel to the $z$-axis toward positive $z$-values.

With reference to Cartesian coordinates $x_f$, $y_f$, and $z_f$, the field propagates toward positive $z_f$-values, has an exponentially decreasing amplitude in the direction of the positive $x_f$-axis, and consists of the transversal components $E_T = E_{x_f}$ and $B_{y_f}$, and the longitudinal component $E_L = E_{z_f}$.

Figure 1 shows a cross-section of the upward moving shell, its velocity $\mathbf{u_S}$ forming the angle $\alpha$ with the phase velocity $\mathbf{u_P}$ of the field. The gray arrows represent the field components $E_L$ and $E_T$. At the thin oblique lines parallel to the $x_f$-axis, $E_T = 0$, and $|E_L|$ is at its maximum. The decreasing amplitude is indicated by the two $E_T$ arrows of different lengths. $F_C$ is the force exerted on the center charge by $E_L$, $F_L$ and $F_T$ are exerted on the shell by $E_L$ resp. by $E_T$. $F_C$ and $F_L$ are not shown in Fig. 1.

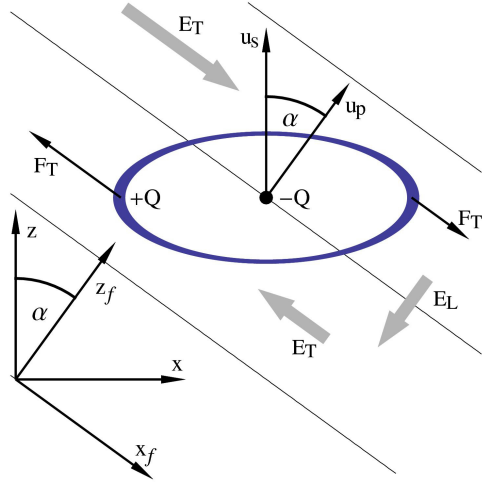

Fig. 1 Cross-section of shell, velocities, electric field components, and force

With $\alpha = 0$, $F_C + F_L = 0$ independently of the phase of $E_L$ at the location of the center charge, as expected. If the center is placed at a point where $E_T = 0$ as in Fig. 1, $F_T = 0$ because the shell is immersed symmetrically into two $E_T$-half-waves of different direction. With the center placed at $E_T = 0$, the $E_L$ component forms a saddle surface at the place of the shell, and $\alpha = 0$ results in the maximum force $F_L$ exerted on the charged spheroid.

When $\alpha$ increases, the extend of the shell along the $x_f$-axis decreases, thereby decreasing $F_L$, and the extend along the $z_f$-axis increases, decreasing $F_L$ further. The shell is now located in two $E_T$-half-waves of different mean strength. For the case $\alpha = \pi/5$, the components of $F_T$ are therefore depicted in Fig. 1 by arrows of different lengths.

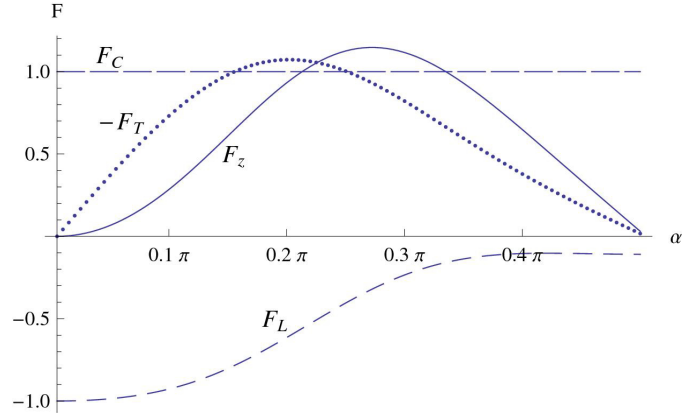

Fig. 2 Forces exerted on the charge as a function of $\alpha$

Figure 2 shows the dependence of the forces on the angle $\alpha$. For each value of $\alpha$, $u_P$ was adjusted to yield a phase-locked movement of the shell, the radius of the shell was set to $\lambda/2$, and the field strength was adjusted to result in $F_C = 1$. $F_L$ and $-F_T$ are shown as computed parallel to the $z_f$-axis resp. antiparallel to the $x_f$-axis. The $z$-component $F_z = (F_C + F_L)\cos\alpha - F_T \sin\alpha$ of the total force $\mathbf{F}_{tot}$ raises even above $F_C$ for $\alpha$ in the range $\pi/4$ to $\pi/3$.

Because $\mathbf{F}_{tot} \cdot \mathbf{u}_S > 0$ for $0 < \alpha < \pi/2$, there is obviously an energy transfer from the incident field to the nonradiating system. The flow of energy from the incident field to the shell as expressed by the Poynting vector is limited to the inside of the shell. From a classical point of view, the system could only absorb the field energy which is currently available inside the shell. In a real experiment, the shell were bound to collide with the material boundary along which the field is propagating after a short time interval, of course.

Beside the force exerted by the external field on the moving shell, there are two other sources or sinks of energy: The growing mechanical stress of the shell and the energy difference created by the superposition of the external and internal fields while the strength of the external field increases. To prove that nevertheless a transfer of energy to or from the remote system takes places, at least three ways can be taken:

- For a given polarity of the charges of the nonradiating system, both the direction of the forces and the sign of the energy difference due to superposition depend on the polarity of the half waves the shell is immersed in but the amount of stress does not. From the assumption that independent of the polarity of the half waves there is no energy transfer to the remote system: $W_F + W_{Su} + W_{St} = 0 = -(W_F + W_{Su}) + W_{St}$ follows
$$W_F + W_{Su} = -(W_F + W_{Su}) = -W_{St} \qquad (5)$$
with the energy $W_F$ associated with the force exerted by the external field, the energy difference $W_{Su}$ caused by superposition, and the energy $W_{St}$ associated with the mechanical stress. However, Eq. 5 can be true only if $W_{St} = 0$ which is certainly not the case.

- The point charge can be replaced by a second charged shell with a diameter less than that of the first one. The diameter of both concentric shells can be varied without causing radiation (as discussed in section B below in more detail). If both diameters are the same at the start and at the end of the observation period then the energy difference $W_{Su}$ is zero at the beginning and at the end. However, the force caused by the $B$ component of the external field is not zero while the diameters are changing. In order to avoid compensation of the forces exerted by the $E$ component and by the $B$ component, the speed of the shells has to differ slightly from the phase velocity of the plane wave. Then the timing of the variation of the diameters in relation to the phase of the external field can be chosen in such a way that, for example, the integral of the $B$ force over time vanishes but the integral of the $E$ force over time does not.

- Instead of letting the charge of the nonradiating system vanish at the beginning and the end of the observation period, one can consider a spatially limited radiation field such as a beam. This is the subject of the following section.

*B. Shell with Variable Radius in a Radially Polarized Beam*

The motivation to investigate the effect of an inhomogeneous field on a charged shell sprang from a similar approach using a beam. With this earlier setup, the effect cannot be easily visualized because the field components of a beam near the focus are more complicated than a plane field, the force caused by the $B$ component cannot be neglected because the radius of the shell varies, and the magnitude of the effect is small. Nevertheless it is included here in order to show that nonradiating absorption is not limited to evanescent waves.

As in the first thought experiment, the absorbing system is a nonconducting, evenly charged spherical shell. Seen from the rest frame of the center of the shell (SRF), it does not radiate if it oscillates radially with an arbitrary frequency spectrum and with both amplitude and phase independent of the angle. Since the power radiated by a charge is the same in all inertial frames the shell remains nonradiating as long as its center moves without acceleration. The mechanism maintaining the oscillation has to overcome the self-force of the charge. After the radius has returned to its initial value the total energy spent is zero, because there is no radiation loss.

Incident radiation fields do not contribute to the energy turnover of the oscillation: Seen from the SRF, the charge is moving exclusively radially in the B component of any external field resulting in the spent power $P_{BR}(t) = 0$. The power $P_{ER}(t)$ exchanged between the E component and the charge $Q$ on the shell S is given by

$$P_{ER}(t) = \oiint_S \sigma_q(t) \mathbf{E}(\mathbf{r},t) \cdot \mathbf{v_R}(t) \, dS = Q \, \mathbf{v_R}(t) \cdot \hat{\mathbf{n}} \oiint_S \mathbf{E}(\mathbf{r},t) \cdot \hat{\mathbf{n}} \, dS = 0 \qquad (6)$$

because $\oiint_S \mathbf{E}(\mathbf{r},t) \cdot \hat{\mathbf{n}} \, dS = 0$ for the incident field $\mathbf{E}(\mathbf{r},t)$, with charge density $\sigma_q(t)$, radial velocity $\mathbf{v_R}(t)$, differential shell area $dS$, and unit vector $\hat{\mathbf{n}}$ normal to $dS$. So, the total power exchanged between the incident field and the mechanism driving the oscillation is $P_R(t) = P_{BR}(t) + P_{ER}(t) = 0$ in every moment. Since the amount of energy spent or gained by a system does not depend on the frame of reference, $P_R(t) = 0$ in every inertial frame. However, seen from frames other than SRF a translational movement of the charge is added, and amplitude and phase of the radial velocity depend on location, due to relativistic effects. Generally, only $P_{ER}(t) = -P_{BR}(t)$ holds.

In the present case, the elimination of the field outside the shell by a center charge is not necessary because the source of the beam can be located so distantly that the quasi-static field of the shell has no effect on it. The shell traverses the focus perpendicular to the axis of the beam; its radius increases and decreases once while passing through the focus; please see the equation for $R'_v(t')$ in the appendix.

Calculations were performed with a beam of frequency $f = 3\,\mathrm{GHz}$ and $k w_0 = 1.3\pi$, and the shell parameters $R_0 = 0.1\lambda$, $u_S = 0.9c$, $\tau = 1/(2\omega)$, $t_0 = T/\gamma$, and $t_1 = 0$, $\lambda$, $\omega$, and $T$ being the wavelength, the angular frequency resp. the period of the beam. The mean force acting on the shell while completely traversing the beam vanishes if $A_S = 0$ (constant radius), as expected, but it does not vanish if $A_S = 0.1\lambda$. A series of shells with thus varying radius crossing the focus with the same frequency as the beam, each shell carrying a charge of $1\,\mathrm{nC}$, equivalent to a current of $3\,\mathrm{A}$, results in a power exchange between the beam and the remote system via the shells of about 400 ppm of the beam's power.

This second thought experiment demonstrates an additional aspect which was perhaps not obvious in the first experiment: While the conservation of energy would be satisfied by a slight attenuation or amplification of the beam behind the focus, the conservation of momentum requires a minor deflection of the beam.

*C. Effect of a UWB field on a system comprising DC current and static H field*

The third thought experiment differs from the preceding ones in at least two aspects:

- The involved macroscopic objects are resting in the laboratory frame.

- Primarily affected is not the radiation but a static field caused by a DC current.

The central question of this thought experiment is: What happens when a unipolar radiation pulse is absorbed while passing through a static magnetic field? The details are outlined in the following points:

a) The set-up consists of a radiator R, a transparent structure M carrying a current with a DC component, and a perfect absorber A. R radiates well separated unipolar pulses of alternating polarity. Due to the shape of M, the static field accompanying the DC current in M is located exclusively on the side of M facing A.

b) The radiation field incident on M does not effectively exchange radiation energy with the sources in M but acts as a "catalyst" by causing an exchange of energy between the current sources and the static field of M. Of course, the integral of this energy, taken over one period of the radiation, amounts to zero.

c) The distance between R and A is less than half the fundamental wavelength of the radiation so there is never more than one unipolar pulse in existence. (At this short distance, the fields of R are a mixture of radiation and reactive components, of course.) For the sake of simplicity, the following description is limited to pulses of one polarity: The E component of the present pulse is aligned parallel to the DC current while the H component of the pulse is oriented anti-parallel to the static field.

d) Due to the orientation of the fields, the sources in M gain energy during the passing of the pulse while the static field looses the same amount of energy by the superposition of the H component of the pulse. This energy deficit is separated from the sources in M by the propagating pulse.

e) When the pulse is absorbed by A the static field has to be transiently attenuated in order to account for the energy deficit caused by the now vanishing superposition. During the subsequent restoration of the static field to its stable state the energy gained before has to be drained from the sources in M. This flow of energy from the sources into the static field has to be accompanied by an electric field.

f) If this electric field would behave according to $\nabla \times \mathbf{E} = -\frac{\partial \mathbf{B}}{\partial t}$ it would cause an increase of energy in M during attenuation of the static H field, and a decrease of energy in M during restauration of the static field resulting in no overall change of energy in M. Consequently, the energy in M gained during the passing of the pulse would remain as excess energy. Obviously this excess energy would vanish during the passing of the subsequent pulse of opposite polar-ity. However, the principle of conservation of energy does not permit even a temporary excess or deficit of energy.

According to the solution proposed here, during the restoration of the static field the behavior of the electric field is indeed governed by $\nabla \times \mathbf{E} = -\frac{\partial \mathbf{B}}{\partial t}$ because $\mathbf{J}_{\varsigma m} = \mathbf{0}$ but during the deflation of the static field the electric field is given by $\nabla \times \mathbf{E} = -\frac{\partial \mathbf{B}}{\partial t} - \mathbf{J}_{\varsigma m}$ with $\mathbf{J}_{\varsigma m} = -\frac{\partial \mathbf{B}}{\partial t}$. As a consequence, the contributions of $\frac{\partial \mathbf{B}}{\partial t}$ and $\mathbf{J}_{\varsigma m}$ cancel each other resulting in $\nabla \times \mathbf{E} = \mathbf{0}$. (A weak analogy is the discharge of a plate capacitor through its slightly conducting dielectric; in that case $\frac{\partial \mathbf{D}}{\partial t}$ and $\mathbf{J}_{\rho e}$ compensate each other resulting in $\nabla \times \mathbf{H} = \mathbf{0}$.) In the present case, the resulting E field drains the correct amount of energy from the sources in M.

When limiting the thought experiment to a specific shape, a cylindrical set-up which extends to infinity in z-dimension seems to provide at least two advantages: If the DC current in the open cylinder M flows parallel to the z-axis then the static H field of M is limited to the outside of M thus satisfying the condition described in point a) provided R is located inside M and A is outside. Additionally, the treatment is reduced to one dimension, $\rho$, if R, M, and A are coaxial.

R forms a line radiator identical to the one described on page 506 of [5]. A harmonic current $I$ with angular frequency $\omega$ and angular wave number $k$ causes the fields

$$Eh_z(\rho,t,\omega) = -\frac{\omega \mu I}{4} H_0^{(2)}(k\,\rho)\,e^{j\omega t} \qquad (7)$$

and
$$Hh_\varphi(\rho,t,\omega) = \frac{\omega I}{4\,c\,j} H_1^{(2)}(k\,\rho)\,e^{j\omega t} \qquad (8)$$

with the Hankel functions $H_i^{(2)} = J_i - j\,Y_i$. An approximately "rectangular" current as the one shown in Fig. 3 a) would yield fields with a temporal evolution as depicted by Fig. 3 b). In order

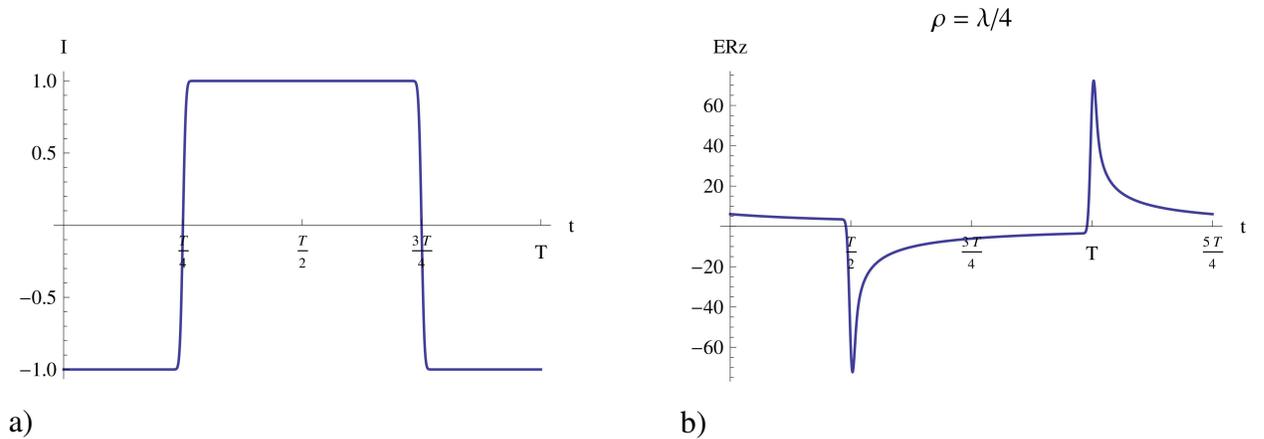

Fig. 3  a) Nearly rectangular current, and b) E field caused by this current to get well separated field pulses, a current described by

$$I(t) = \frac{1}{n} \sum_{i=1}^{n-1} \operatorname{Re}\left( F_i\, e^{ji\omega_0 t}\, e^{j\frac{\pi}{4}} \sqrt{\frac{i\,\omega_0}{c}} \right) \tag{9}$$

is employed, with the Fourier coefficients $F_i$ of the current of Fig. 1 a), and the fundamental angular frequency $\omega_0$. Then the field components are given by

$$E_{Rz}(\rho,t) = \frac{1}{n} \sum_{i=1}^{n-1} \operatorname{Re}\left( F_i\, Eh_z(\rho,t,i\,\omega_0)\, e^{j\frac{\pi}{4}} \sqrt{\frac{i\,\omega_0}{c}} \right) \tag{10}$$

and
$$H_{R\varphi}(\rho,t) = \frac{1}{n} \sum_{i=1}^{n-1} \operatorname{Re}\left( F_i\, Hh_\varphi(\rho,t,i\,\omega_0)\, e^{j\frac{\pi}{4}} \sqrt{\frac{i\,\omega_0}{c}} \right) \tag{11}$$

where $Eh_z$ and $Hh_\varphi$ are the same as described by (7) and (8). Fig. 4 shows the current and the field $E_{Rz}$. It should be noted that both $E_{Rz}$ and $H_{R\varphi}$ do not exactly vanish between the pulses. As a consequence, there is a small centripetal flow of energy, and the energy in M varies slightly between the pulses.

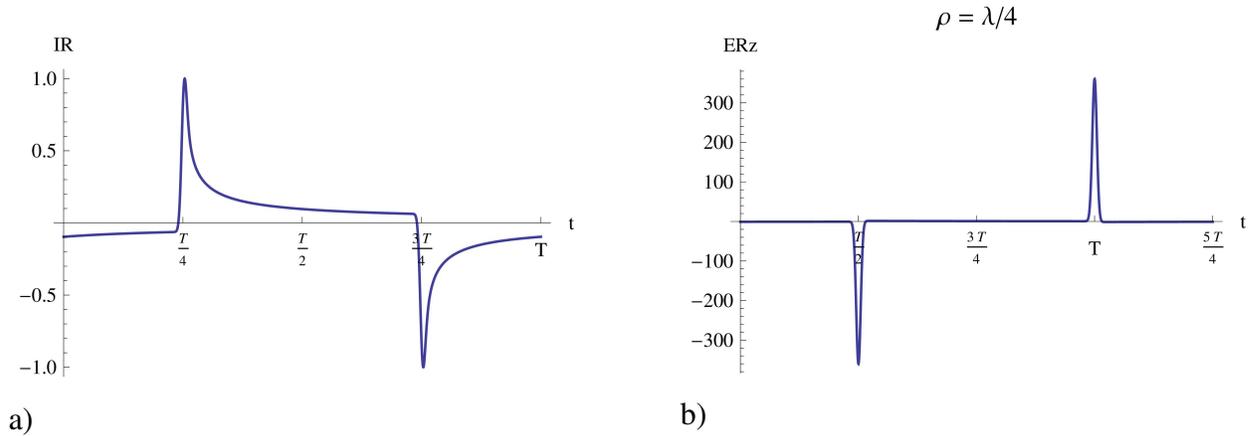

Fig. 4 a) Current according to (9), and b) E field as described by (10)

The thin-walled cylinder M is completely transparent to the fields of R, and carries a DC current parallel to the $z$-axis. (Since "DC" and "AC" are commonly used to denote quantities with a frequency $f = 0$ resp. $f \neq 0$, I'm writing "DC current" or "AC current" despite the fact that the "C" stands for "current".) In order to achieve both properties, it is made from electrically very short pieces of a PEC connected by current sources which deliver a constant DC current superimposed on an AC current. The AC current renders M exactly non-radiating despite the incident field by R.

It is worth noting that the pieces of PEC in M can be made so small that the electrical multipole field caused by the sources does virtually vanish at $\rho = 0$. Since M does not radiate it has virtually no effect on R. The static field caused by the DC current $I_M$ in M is zero inside M and given by $H_M(\rho) = \dfrac{I_M}{2\pi\rho}$ on the outside. As a consequence, the power flow inside M caused by R is not affected by M.

The thick-walled cylinder A acts as an absorber of the radiation by R. Absorbers based on a waveguide approach can be made as thin as $\dfrac{\lambda}{10}$ [6]. In the following example, complete absorption is assumed but this is not a crucial feature.

The following is meant to illustrate the paradox. The current in R corresponds to Fig. 4 a) with an amplitude of 1 A and a fundamental frequency $f_0 = 67.8$ MHz resulting in $\lambda_0 \approx 4.42$ m for the fields. The pulse width $pw$ of the fields (10% limit) is about 130 mm. The $F_i$ were calculated by applying FFT on 1000 points per period, one quarter of the resulting coefficients were used, $n = 250$. The radius of M is $r_M = \dfrac{\lambda}{4}$, the inner radius of A is $r_{Ai} = 1.6$ m, the outer radius $r_{Ao} = 2$ m $< \dfrac{\lambda}{2} - pw$. As a consequence of the value of $r_{Ao}$, there exists at most one pulse inside the setup. The DC current in M is $I_M = 10$ A. Fig. 5 on the following page shows in 8 temporal stages the energies resp. the energy densities between $\rho = \dfrac{13}{64}\lambda$ and $\rho = \dfrac{\lambda}{2}$ for a length of the setup in $z$-dimension of 1 m. The first value in the captions gives the field energy between $\rho = \dfrac{13}{64}\lambda$ and M, and is calculated by

$$W_i(t) = 2\pi \int_{\frac{13}{64}\lambda}^{\frac{\lambda}{4}} \rho \left( \dfrac{\varepsilon_0}{2} E_{Rz}^2(\rho,t) + \dfrac{\mu_0}{2} H_{R\varphi}^2(\rho,t) \right) d\rho \qquad (12)$$

The change of energy in the sources of M since the start of observation is the second value:

$$W_M(t) = I_M \int_{to}^{t} E_{Rz}\left(\dfrac{\lambda}{4}, \tau\right) d\tau \qquad (13)$$

with start of the observation $to$. In the graphics, $W_M$ is depicted as a line in order to render it more visible; however, the thickness of M is assumed to be very small. The third value in the captions gives the field energy outside M:

$$W_o(t) = 2\pi \left( \int_{\frac{\lambda}{4}}^{2} \rho \left( \dfrac{\varepsilon_0}{2} E_{Rz}^2(\rho,t) + \dfrac{\mu_0}{2} \left( H_{R\varphi}(\rho,t) + H_M(\rho) \right)^2 \right) d\rho + \int_{2}^{\frac{\lambda}{2}} \rho \dfrac{\mu_0}{2} H_M^2(\rho) d\rho \right) \qquad (14)$$

The energy converted by A into heat since the start of observation is calculated by

$$W_A(t) = 4\pi \int_{to}^{t} -E_{Rz}(2,\tau) H_{R\varphi}(2,\tau) \, d\tau \tag{15}$$

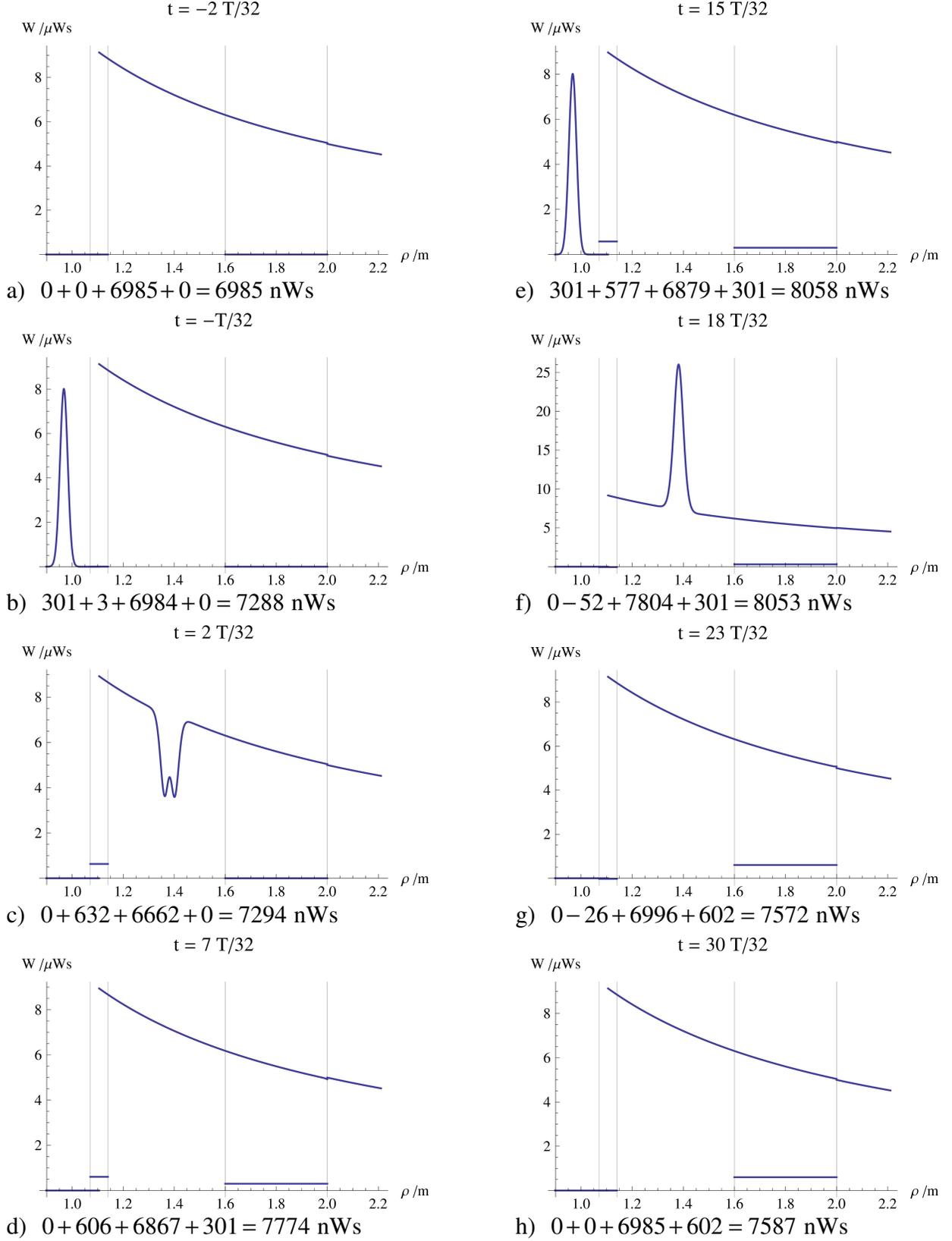

Fig. 5 Energies resp. energy densities during one period $T_0$

The usage of the Poynting vector of the <u>pulse</u> in (15) assumes that the absorber comprises only materials with $\mu_r = 1$, and effectively reacts only on $E_{Rz}$. $E_{Rz}$ causes currents in A, and these currents are accompanied by a curl of $H$ which in turn causes power flows into the dissipative elements of A. The static field $H_M$ does not affect the function of A. This (together with the spatial limitation to one pulse at a time) seems to be the crucial point of the paradox. In the captions of Fig. 5, the units of the summands (always nWs) are omitted due to a lack of space.

Fig. 5 a) shows the state at the start of observation, $to = -\frac{2}{32} T_0$. In Fig. 5 b), a pulse with $E_{Rz} > 0$ and $H_{R\varphi} < 0$ has entered the observed volume of space, and introduced an additional energy of about 301 nWs. By passing M, the pulse has enhanced the energy in M by about 632 nWs, and reduced the field energy outside M by about 322 nWs as shown in Fig. 5 c). So far the energy is well balanced (as mentioned above, slight changes of energy are caused by the small but non-zero values of $E_{Rz}$ and $H_{R\varphi}$ between the pulses).

Fig. 5 d) depicts the paradox: M has retained about 606 nWs, during the absorption of the pulse the field energy of $H_M$ has regained (nearly) its original value, and the heat energy in A corresponds exactly to the energy of the pulse. As a consequence, the energy in M is excess energy. The rest of Fig. 5 shows that at the end of one period, $t = to + T_0$, the energy is balanced again: The energy of $H_M$ is exactly restored to its original value, and the heat energy in A equals the energy of two pulses. But this does not lessen the paradox: The principle of conservation of energy has to be valid at all times.

Even if the solution provided here solves the paradox, several questions remain to be answered by real experiments: Does the deflation or inflation of the static field spread out from the location of absorption to the whole static field, and if so, at which speed? Does the deflation or inflation remain strictly limited to the static field or does it affect the radiation fields as well?

## IV. DISCUSSION

*Immediate Impact on the Theory*

All three thought experiments result in a paradox: In the first and the second one, while the incident fields exert a force parallel to the velocity of the shell and therefore transfer energy and momentum via the shell to the remote system, Eq. 3 and 4 force the incident fields to continue unchanged in the absence of a secondary field emitted by the shell. In the third one, unipolar radiation pulses displace energy of a non-radiating system in such a way that the absorption of the pulses causes a temporary excess or deficit of the system's energy. Deflating fields would lose energy and momentum, inflating fields would gain both, so no conservation law would be violated. In order to enable the deflation or inflation of radiation fields, the densities of particle-free divergence $\varsigma_e$ and $\varsigma_m$ have to be nonzero. The virtual velocities $\mathbf{v_e}$ and $\mathbf{v_m}$ in Eq. 1 and 2 can be determined by using these same equations. Setting the terms $\rho_e \mathbf{u_e}$ and $\rho_m \mathbf{u_m}$ to zero, and replacing $\varsigma_e$ and $\varsigma_m$ by $\nabla \cdot \mathbf{D}$ resp. $\nabla \cdot \mathbf{B}$ results in

$$\mathbf{v_e} = (\nabla \times \mathbf{H} - \partial \mathbf{D}/\partial t) / \nabla \cdot \mathbf{D} \tag{16}$$

$$\mathbf{v_m} = -(\nabla \times \mathbf{E} + \partial \mathbf{B}/\partial t) / \nabla \cdot \mathbf{B} \tag{17}$$

Magnitude and direction of $\mathbf{v_e}$ and $\mathbf{v_m}$ depend on the orientation of the gradient of attenuation relative to both the direction of propagation and the orientation of $\mathbf{E}$ resp. $\mathbf{H}$. As an example, consider a homogeneous plane wave consisting of an $E_x$-component and a $H_y$-component, propagating parallel to the $z$-axis toward positive $z$-values. A gradient of attenuation independent of location, and given by the elevation angle $\vartheta$ and the azimuthal angle $\varphi$ results in

$$\mathbf{v_e} = -c \cot \vartheta \sec \varphi \, \mathbf{e_x} + 0 \, \mathbf{e_y} + c \, \mathbf{e_z} \tag{18}$$

$$\mathbf{v_m} = 0 \, \mathbf{e_x} - c \cot \vartheta \csc \varphi \, \mathbf{e_y} + c \, \mathbf{e_z} \tag{19}$$

with the standard basis vectors $\mathbf{e_x}$, $\mathbf{e_y}$, and $\mathbf{e_z}$. Obviously, $c \leq |\mathbf{v_e}| \leq \infty$ and $c \leq |\mathbf{v_m}| \leq \infty$. This poses no problem because these virtual velocities are not associated with the motion of matter or energy. Put in a pictorial way, if Faraday's "lines of force" would propagate with the radiation field, $\varsigma_e \neq 0$ and $\varsigma_m \neq 0$ would indicate that some of these lines have "open ends", with $\mathbf{v_e}$ and $\mathbf{v_m}$ describing the movement of these open ends.

With the current densities $\mathbf{J_{\rho e}} = \rho_e \mathbf{u_e}$, $\mathbf{J_{\varsigma e}} = \varsigma_e \mathbf{v_e}$, $\mathbf{J_{\varsigma m}} = \varsigma_m \mathbf{v_m}$, and $\rho_m = 0$ confirming the absence of magnetic monopoles, the Maxwell equations in the usual redundant form read

$$\nabla \times \mathbf{E} = -\partial \mathbf{B}/\partial t - \mathbf{J_{\rho m}} - \mathbf{J_{\varsigma m}} \tag{20}$$

$$\nabla \times \mathbf{H} = \partial \mathbf{D}/\partial t + \mathbf{J_{\rho e}} + \mathbf{J_{\varsigma e}} \tag{21}$$

$$\nabla \cdot \mathbf{D} = \rho_e + \varsigma_e \tag{22}$$

$$\nabla \cdot \mathbf{B} = \varsigma_m \tag{23}$$

supplemented by the continuity equations $\nabla \cdot \mathbf{J_{\rho e}} = -\partial \rho_e / \partial t$, $\nabla \cdot \mathbf{J_{\varsigma e}} = -\partial \varsigma_e / \partial t$, $\nabla \cdot \mathbf{J_{\rho m}} = 0$, and $\nabla \cdot \mathbf{J_{\varsigma m}} = -\partial \varsigma_m / \partial t$. $\mathbf{J_{\rho m}}$ may be caused by orientation-changing magnetic dipoles [7].

*Verification and Application*

Of course, the enhanced Maxwell equations have to be supplemented by further modifications of the theory, and several questions have to be answered, first of all, if and how the existence of "deflating" or "inflating" radiation fields could be proved or disproved experimentally.

In the framework of the current technology, it is not possible to implement the first two thought experiments in reality. However, since the total effect of a set of charges is the linear superposition of the effects of all the single charges involved, one could determine a small subset of charges whose effect is maximal, and replace these charges by current pulses on some kind of "one-dimensional active metamaterial". Modulating a sequence of current pulses using a low frequency, and in turn modulating monochromatic radiation by these pulses would result in easily detectable sidebands which could not be generated by superposition.

But with one modification, the third thought experiment seems to provide the easiest implementation: The length of the setup has to be reduced from infinity to a size fitting in common laboratories. This can be achieved by terminating the ends of the radiator, the modulator, and the absorber by horizontal disks which act as mirrors as well as waveguide. The return path of the modulator is located outside of A; thus, the modulator is actually the inner part of a toroidal coil.

By comparing the signals of two different kind of magnetic field sensors, one kind based on electrically conducting loops and the other on the magneto-optical effect, it should be possible to verify the "deflation" and "inflation" of the static H field not accompanied by an electric field.

After experimental confirmation, the question of applications would arise. By multiplicatively attenuating radiation fields, low frequency components in UWB fields can be generated. This will enable the stimulation of neurons for diagnostic and therapeutic purposes, using UWB fields focused to a smaller volume than is possible by employing near fields as is the state of the art. Advancing into the mm-wave domain, the presentation of surfaces with software-controlled virtual tactile properties (Braille, texture) might be feasable by stimulation of skin receptors.

## APPENDIX: EQUATIONS USED IN THE CALCULATIONS

Two rest frames are used, one defined by a fictitious laboratory (LRF) which is identical to the rest frame of the focus in the second thought experiment, the other by the center of the spherical shell (SRF). Quantities as seen from SRF are indicated by an apostrophe ('). Spatial data refer to Cartesian coordinates. Since movements are mostly directed along the $z$-axis, the $z$-axis and the $z'$-axis coincide. The time is so adjusted that $z = 0 \equiv z' = 0$ at $t = t' = 0$, resulting in the relations $x' = x$, $y' = y$, $z' = \gamma(z - u_S t)$, and $t' = \gamma(t - u_S z / c^2)$ with $\gamma = 1/\sqrt{1 - u_S^2/c^2}$ and the relative speed of the rest frames $u_S$. Seen from SRF, the radius of the shell is given by $R'(t') = R'_0 + R'_v(t')$ with $R'_v(t') = A_S \left( erf((t' + t_0)/\tau) - erf((t' + t_1)/\tau) \right)$, and $R'_0$, $A_S$, $t_0$, $t_1$, and $\tau$ being constants. For the radial velocity follows $u'_R(t') = 2 A_S \left( \exp(-(t' + t_0)^2/\tau^2) - \exp(-(t' + t_1)/\tau^2) \right)/(\tau \sqrt{\pi})$. A cross-section of the shell in a plane perpendicular to the $z'$-axis forms a circle with the radius $r'(z', t') = \sqrt{R'(t')^2 - z'^2}$ and seen from LRF $r(z, t) = r'(z'(z, t), t'(z, t))$.

With reference to SRF, the charge density is $\sigma'_e(t') = Q/(4\pi R'(t')^2)$ with the total charge $Q$. With a shell of constant radius ($A_S = 0$), the charge density as seen from LRF is $\sigma_e(z) = Q\sqrt{\cos^2(\vartheta) + (\sin(\vartheta)/\gamma)^2}/(4\pi R_0^2)$ with $\vartheta = \cos^{-1}(\gamma z / R_0)$ denoting the elevation angle with respect to the $z$-axis. If the incident radiation is independent of the y-coordinate, the $z$-component of the electric field acting on the charge at the plane defined by $z = \zeta$ when the center of the shell is at $(x, z, t)$ is given by

$$E_\zeta(x, z, \zeta, t) = \int_0^{2\pi} \text{Re}\left( E_z\left(x + R_0 \sqrt{1 - (\gamma(z-\zeta)/R_0)^2} \cos\varphi, \zeta, t\right) \right) d\varphi.$$ The total force acting in $z$-direction on the shell is obtained by integrating over the $z$-axis:

$$F_z(x,z,t) = R_0 \int_{z-\frac{R_0}{\gamma}}^{z+\frac{R_0}{\gamma}} E_\zeta(x,z,\zeta,t)\,\sigma_e(\zeta)\,\sqrt{1 - R_0(\gamma^4 - \gamma^2)\zeta^2}\; d\zeta .$$

If the radius of the shell is time-dependent, the charge density is not only affected by relativistic length contraction but also by the deformation of the shell due to the dependence of the phase of the radial oscillation on the $z$-coordinate, as seen from LRF. In order to avoid computing the charge density explicitly, the forces are obtained by integrating the charge density belonging to $R'$ instead of $R$ over $\vartheta'$ but multiplying with the incident field at the position where the charge is seen from LRF. For a given value of $\vartheta'$ the value of $z$ with $\vartheta'(z,t) = \cos^{-1}(z'(z,t)/R'(t'(z,t)))$ is numerically approximated. Then $r(z,t)$, $u_R'(t'(z,t))$, and $u_R = u_R' \sin\vartheta' / (\gamma(1 + u_s\, u_R' \cos(\vartheta')/c^2))$ can be determined. The forces acting on the charged circle defined by $\vartheta'$ are integrated,

$$F_{CEz}(\vartheta',t) = \frac{Q\,r}{4\pi R'(t'(z,t))} \int_0^{2\pi} \operatorname{Re}\!\left(E_z(x_{\mathit{off}} + r\cos\varphi, r\sin\varphi, z, t)\right) d\varphi \quad \text{and}$$

$$F_{CBz}(\vartheta',t) =$$

$$\frac{Q\,r\,u_R}{4\pi R'(t'(z,t))} \int_0^{2\pi} \operatorname{Re}\!\left(-B_x(x_{\mathit{off}} + r\cos\varphi, r\sin\varphi, z, t)\sin\varphi + B_y(x_{\mathit{off}} + r\cos\varphi, r\sin\varphi, z, t)\cos\varphi\right) d\varphi$$

($R'$ in the denominator in both $F_{CEz}$ and $F_{CBz}$ is not squared because otherwise it had to be multiplied in the integration over $\vartheta'$.) These forces are integrated over $\vartheta'$:

$$F_{Ez}(t) = \int_0^\pi F_{CEz}(\vartheta',t)\,d\vartheta' \quad \text{and} \quad F_{Bz}(t) = \int_0^\pi F_{CBz}(\vartheta',t)\,d\vartheta' .$$

Please note that both sets of field equations comply with the Maxwell equations identically. The first set describes the inhomogeneous plane field:

$$x_f = x\cos\alpha - z\sin\alpha, \qquad z_f = x\sin\alpha + z\cos\alpha, \qquad E_T = A_f \cosh(\beta)\,e^{-\sinh(\beta)k\,x_f}\,e^{i(\omega t - \cosh(\beta)k\,z_f + \psi)},$$

$$E_L = -A_f \sinh(\beta)\,e^{-\sinh(\beta)k\,x_f}\,e^{i(\omega t - \cosh(\beta)k\,z_f + \psi)}, \qquad E_x = E_T \cos\alpha + E_L \sin\alpha, \qquad E_y = 0,$$

$$E_z = -E_T \sin\alpha + E_L \cos\alpha,\; B_x = 0,\; B_y = \frac{A_f}{c}e^{-\sinh(\beta)k\,x_f}\,e^{i(\omega t - \cosh(\beta)k\,z_f + \psi)},\; B_z = 0.$$

The equations of the radially polarized beam are obtained by a method following Mitri [8]. Deviating from Mitri's procedure, here the beam is limited to asymmetric components, resulting in $H_z = 0$. The field components are derived from a vector potential field with only a $z$-component

$$A_z = A_0 \frac{zr}{2\sinh^2(k\,zr)}\left(\frac{\sin(k\,R^-)}{R^-}e^{k\,zr} - \frac{\sin(k\,R^+)}{R^+}e^{-k\,zr}\right) \quad \text{with} \quad R^\pm = \sqrt{x^2 + y^2 + (z \pm i\,zr)^2}\,, \text{ according}$$

to $\mathbf{H} = \frac{1}{\mu}\nabla\times\mathbf{A}$ and $\mathbf{E} = \frac{i}{\varepsilon\omega}\nabla\times\mathbf{H}$. This results in $E_x = \frac{i A_0 c^2 zr}{2\omega \sinh^2(k\,zr)}e^{-i(\omega t+\psi)} f_{Ex}$,

$$E_y = \frac{i A_0 c^2 zr}{2\omega \sinh^2(k\,zr)}e^{-i(\omega t+\psi)} f_{Ey},\quad E_z = \frac{i A_0 c^2 zr}{2\omega \sinh^2(k\,zr)}e^{-k\,zr - i(\omega t+\psi)} f_{Ez},$$

$$H_x = \frac{A_0\,zr}{2\mu \sinh^2(k\,zr)}e^{-i(\omega t+\psi)} f_{Hx}, \text{ and } H_y = \frac{A_0\,zr}{2\mu \sinh^2(k\,zr)}e^{-i(\omega t+\psi)} f_{Hy}$$

with

$$f_{Ex} = -\frac{3e^{kzr}\,k\,x(z-i\,zr)\cos(k\,R^-)}{(R^-)^4} + \frac{3e^{-kzr}\,k\,x(z+i\,zr)\cos(k\,R^+)}{(R^+)^4} + \frac{3e^{kzr}\,k\,x(z-i\,zr)\sin(k\,R^-)}{(R^-)^5} -$$

$$\frac{e^{kzr}\,k^2\,x(z-i\,zr)\sin(k\,R^-)}{(R^-)^3} - \frac{3e^{-kzr}\,x(z+i\,zr)\sin(k\,R^+)}{(R^+)^5} + \frac{e^{-kzr}\,k^2\,x(z+i\,zr)\sin(k\,R^+)}{(R^+)^3}$$

$$f_{Ey} = -\frac{3e^{kzr}\,k\,y(z-i\,zr)\cos(k\,R^-)}{(R^-)^4} + \frac{3e^{-kzr}\,k\,y(z+i\,zr)\cos(k\,R^+)}{(R^+)^4} + \frac{3e^{kzr}\,k\,y(z-i\,zr)\sin(k\,R^-)}{(R^-)^5} -$$

$$\frac{e^{kzr}\,k^2\,y(z-i\,zr)\sin(k\,R^-)}{(R^-)^3} - \frac{3e^{-kzr}\,y(z+i\,zr)\sin(k\,R^+)}{(R^+)^5} + \frac{e^{-kzr}\,k^2\,y(z+i\,zr)\sin(k\,R^+)}{(R^+)^3}$$

$$f_{Ez} = \frac{e^{2kzr}\,k\,y(x^2+y^2-2(z-i\,zr)^2)\cos(k\,R^-)}{(R^-)^4} - \frac{k(x^2+y^2-2(z+i\,zr)^2)\cos(k\,R^+)}{(R^+)^4} - \frac{3e^{2kzr}\,x^2\sin(k\,R^-)}{(R^-)^5} -$$

$$\frac{3e^{2kzr}\,y^2\sin(k\,R^-)}{(R^-)^5} + \frac{2e^{-2kzr}\sin(k\,R^-)}{(R^-)^3} + \frac{e^{2kzr}\,k^2\,x^2\sin(k\,R^-)}{(R^-)^3} + \frac{e^{2kzr}\,k^2\,y^2\sin(k\,R^-)}{(R^-)^3} +$$

$$\frac{3x^2\sin(k\,R^+)}{(R^+)^5} + \frac{3y^2\sin(k\,R^+)}{(R^+)^5} - \frac{2\sin(k\,R^+)}{(R^+)^3} - \frac{k^2\,x^2\sin(k\,R^+)}{(R^+)^3} - \frac{k^2\,y^2\sin(k\,R^+)}{(R^+)^3}$$

$$f_{Hx} = \frac{e^{kzr}\,k\,y\cos(k\,R^-)}{(R^-)^2} - \frac{e^{-kzr}\,k\,y\cos(k\,R^+)}{(R^+)^2} - \frac{e^{kzr}\,y\sin(k\,R^-)}{(R^-)^3} + \frac{e^{-kzr}\,y\sin(k\,R^-)}{(R^+)^3}$$

$$f_{Hy} = \frac{e^{kzr}\,k\,x\cos(k\,R^-)}{(R^-)^2} - \frac{e^{-kzr}\,k\,x\cos(k\,R^+)}{(R^+)^2} - \frac{e^{kzr}\,x\sin(k\,R^-)}{(R^-)^3} + \frac{e^{-kzr}\,x\sin(k\,R^-)}{(R^+)^3}$$